# Influences of interfacial oxidization on surface magnetic energy, magnetic damping and spin-orbit-torques in Pt / ferromagnet / capping structures

D. J. Lee[1,2], W. M. Jeong[2,3], D. H. Yun[2,4], S. Y. Park[5], B.-K. Ju[4], K.-J. Lee[1,3], H. C. Koo[1,2],

B.-C. Min[2,6], and O. J. Lee[2*]

[1]KU-KIST Graduate School of Converging Science and Technology, Korea University, Seoul 02841, Korea

[2]Center for Spintronics, Korea Institute of Science and Technology, Seoul 02792, Korea

[3]Department of Materials Science and Engineering, Korea University, Seoul 02841, Korea

[4]Department of Electrical Engineering, Korea University, Seoul 02841, Korea

[5]Spin Engineering Physics Team, Korea Basic Science Institute, Daejeon 34133, Korea

[6]Division of Nano and Information Technology, KIST School, Korea University of Science and Technology, Seoul 02792, Korea

We investigate the effect of capping layer (CAP) on the interfacial magnetic anisotropy energy density ($K_S$), magnetic damping (α), and spin-orbit torques (SOTs) in heavy-metal (Pt) / ferromagnet (Co or Py) / CAP (MgO/Ta, HfO$_x$, or TaN). At room temperature (RT) the CAP materials influence the effective magnitude of $K_S$, which is associated with a formation of interfacial magnetic oxides. The dynamical dissipation parameters of Co are considerably influenced by the CAP (especially MgO) while those of Py are not. This is possibly due to an extra magnetic damping via spin-pumping process across the Co/CoO interface and incoherent magnon generation (spin fluctuation) in the interfacial CoO. It is also observed that both anti-damping and field-like SOT efficiencies vary marginally with the CAP in the thickness ranges we examined. Our results reveal the crucial role of interfacial oxides on the perpendicular magnetic anisotropy, magnetic damping, and SOTs.

---

* ojlee@kist.re.kr

Electrical manipulation of magnetization[1,2,3,4,5] is of great interest because of the prospective application in low-power and high-speed spintronic devices, as well as the notable scientific advancement. Recently, several works have demonstrated that SOTs[2,3,4,5] induced by in-plane charge currents can efficiently reverse the magnetization in multilayers of heavy-metal(HM)/ferromagnet(FM)/insulator-CAP with strong spin-orbit coupling (SOC)[6,7] (see Fig. 1a). The magnitudes of SOTs are generally characterized by measurable figure of merits that correspond to damping-like (DL, $\tau_{DL}$) and field-like (FL, $\tau_{FL}$) SOT-efficiencies, i.e., $\theta_{DL}$ and $\theta_{FL}$, respectively. The dimensionless $\theta_{DL}$ is also referred to as a spin-Hall angle[2,4] that defines the conversion ratio from charge to spin currents in the HM.

Although most of the research to date has focused on the HM/FM interface at which the generation of non-equilibrium spin-polarization[2,4,7] or transmission[8,9] of spin-current ($J_S$) and the enhancement[10] in α through the spin-pumping process occur, the results of several recent works[11,12,13,14] have suggested that the physical and chemical properties of the CAP can also change the characteristics of spin torques via modification of reflection, absorption, and/or scattering of the $J_S$ at the FM/CAP interface. It is also expected that the different enthalpy of formation of the materials will dissimilarly promote an intermixing (stoichiometry) and an interfacial formation of magnetic oxides, which may change the magnetic and spintronic properties. For instance, Ref. [11] investigated the process by which the oxide interface modifies the $\tau_{FL}$ via the insertion of an ultrathin oxidized Hf layer at the FeCoB/MgO interface in perpendicularly magnetized HM/FeCoB/MgO heterostructures. Our previous work[15] demonstrated that the insertion of an ultrathin magnetic dusting layer between FM and MgO layers can play an important role in the determination of α. Nevertheless, there is still a dearth of experimental work studying how the FM/CAP interface contributes to $K_S$, α, $\theta_{DL}$, and $\theta_{FL}$ in HM/FM/CAP heterostructures.

In this work, we study the effects of interfacial magnetic oxide formation on the magnetic and spintronic properties by examining six different series of layer stacks consisting of Pt/FM/CAP (Fig. 1a). The FM is either Co or Py (=$Ni_{79}Fe_{21}$) and the CAP is insulating MgO/Ta, $HfO_x$ or metallic TaN. As illustrated in Fig. 1(b), the presence of interfacial oxide can give rise to a strong modification of the interfacial magnetic-anisotropy-energy density, magnetic damping, and SOTs. We find, by utilizing X-ray photoelectron spectroscopy (XPS) and spin-torque ferromagnetic resonance (ST-FMR) measurements, that (i) the effective $K_S$ becomes deteriorated with a substantial formation of interfacial magnetic oxides; $K_S$(MgO-CAP) < $K_S$($HfO_x$-CAP) < $K_S$(TaN-CAP) for both Co and Py samples. (ii) The α(Co) is significantly influenced by the CAP whereas the α(Py) is relatively unaffected by the CAP material; α(Pt/Co/$HfO_x$) ≈ α (Pt/Co/TaN) << α(Pt/Co/MgO) whereas α(Pt/Py/MgO) ≈ α(Pt/Py/$HfO_x$) ≈ α (Pt/Py/TaN). (iii) Both the $θ_{DL}$ and $θ_{FL}$ are weakly dependent on the CAP materials. Our results confirm that, though the SOTs originate primarily from the Pt/FM interface, the perpendicular magnetic anisotropy (PMA) and the magnetic damping are significantly influenced by the interfacial formation of *antiferromagnetic* (AF) oxides.

Six series of layer stacks were prepared on thermally oxidized Si substrates (Si/$SiO_x$) using dc/rf magnetron sputtering from two-inch planar targets at room temperature (RT). The structure consists of substrate/Pt(5)/FM($t_{FM}$)/CAP, where the FM is either Co or Py and the CAP is either MgO(2)/Ta(2), $HfO_x$(3) or TaN(3) (nominal thicknesses in nm). The MgO layer is used as a standard tunnel barrier in MRAM technology because it facilitates very effective electrical readout of magnetic change. Therefore, it is necessary to investigate its influence for a given HM/FM bilayer. The $HfO_x$-CAP is an alternative oxide for comparison to MgO-CAP (MgO/Ta). The TaN-CAP is tested as a representative non-oxide because it is stable, metallic but also highly resistive. The thickness of the Co ($t_{Co}$) layer is varied from 3 nm to 15 nm and that of the Py layer ($t_{Py}$) is varied from 2 nm to 10 nm. All of the sputtered materials were deposited on the substrate with an oblique orientation except for the MgO target that was faced towards it. The base pressure of the sputtering chamber was lower than $5 \times 10^{-8}$ Torr, and the deposition rates were lower than 0.5 Å/s.

For the ST-FMR measurement, optical lithography and ion-milling were used to pattern multilayer films onto rectangular shaped bars with 15 μm width ($w$) and 50 μm length ($l$). In a subsequent process step, a waveguide contact made of Ti (10 nm)/Au (100 nm) was defined on top of the samples to facilitate the passage of a RF current through the devices. No post-annealing was carried out as the temperature ($T$) of the samples was kept lower than 110 ºC during the fabrication process. It should be noted that the Ta(2) protective CAP is expected to be fully air-oxidized, so that both MgO(2)/Ta(2) and HfO$_x$(3) are considered as insulators, i.e., no charge and spin currents flow through the layers (see Fig. 1a). The TaN(3) layer is metallic but since its resistivity is at least one or two orders of magnitude higher than that of Pt and FM (≈ 15-30 μΩ-cm), the current shunting through the TaN layer is negligible in the ST-FMR analysis, which will be discussed in the following section. All the measurements were done at RT except for the $T$-dependence of hysteresis loops.

The formation of the interfacial FM-oxide, which are dependent on the used CAP, was identified using XPS on un-patterned films. In-situ ion-etching was used to remove most of the CAPs before investigating the composition of magnetic layers. Figure 2a shows the Co $2p_{3/2}$ XPS spectral region from the Pt/Co(5)/CAP samples. All of the spectra exhibit primary peaks at ~778 eV, confirming the existence of metallic Co. The secondary peak at ~780 eV corresponding to CoO $2p_{3/2}$ shows a variation in the amplitude depending on the CAP materials. The film with the MgO-CAP has the largest peak at 780 eV, indicating the presence of a significant CoO at the Co/MgO interface; the film with the HfO$_x$-CAP exhibits a smaller peak, revealing less formation of interfacial CoO; the sample with the TaN-CAP shows no distinguishable peak. The Ni $2p_{3/2}$ XPS spectral from the Pt/Py(5)/CAP films (Fig.

2b) show primary and satellite peaks at ~853 eV and ~860 eV, respectively, both of which originate from the metallic Ni. A secondary peak at ~855 eV corresponding to NiO is observed only with the MgO-CAP which demonstrates the existence of interfacial NiO. No noticeable oxide peaks are observed for the $HfO_x$-CAP and the TaN-CAP. The Fe $2p_{3/2}$ XPS spectra also exhibited a very consistent dependence of XPS spectra on the CAP (not shown).

The XPS results confirm the presence of the interfacial FM-oxide state, in descending order of oxidation, in the MgO-, $HfO_x$-, and TaN-CAP. A comparison of the relative amplitudes of the secondary peaks between Pt/Co/MgO/Ta and Pt/Py/MgO/Ta films indicates that the degree of oxidation of Py is less than that of Co. This is possibly due to the difference of the enthalpy of formation of oxides with each metallic element (Hf, Mg, Co, Ni, and Fe). For instance, $HfO_2$ is a more stable oxide than MgO, since the enthalpy of formation of $HfO_2$ ($\approx -1120$ kJ/mol) is much larger in magnitude than that of MgO ($\approx -600$ kJ/mol). The formation of the interfacial FM-oxide is relatively easier with the MgO-CAP than the $HfO_x$-CAP. The TaN-CAP protects the FM surfaces from the oxidation as expected.

We examined the effect of CAP on the magnetic properties of Pt/FM($t_{FM}$)/CAP. The saturation magnetization ($M_S$) and magnetic dead-layer thickness ($t^d$) were characterized for the six systems using a vibrating sample magnetometer (VSM). The effective thickness of FM is defined as $t_{FM}^{eff} = t_{FM} - t_{FM}^d$. Figures 3a–b show the measured moment (per unit area) of the un-patterned films as a function of $t_{Co}$ or $t_{Py}$. The linear fits provide the ($M_S$, $t^d$) of Co ≈ (1200 emu/cc, 0.4 nm), (1400 emu/cc, 0.3 nm) and (1430 emu/cc, 0.7 nm) for the MgO-, $HfO_x$- and TaN-CAP respectively. In addition, the ($M_S$, $t^d$) of Py ≈ (870 emu/cc, 1.0 nm), (650 emu/cc, 0.0 nm) and (760 emu/cc, 0.5 nm) for the MgO-, $HfO_x$- and TaN-CAP individually (see Table. 1). Note that the $M_S$ of Co in the Pt/Co/MgO/Ta layer is ~15 % smaller than those in the Pt/Co/$HfO_x$, and Pt/Co/TaN layers. This might be due to a significant Co-oxidation in Pt/Co/MgO/Ta, as revealed by our XPS analysis, or to the diffusion of Mg, Oxygen or Ta into the Co bulk[16]. The CAP material influences the $M_S$ of Py as well.

Next, we systematically measured the frequency-dependent ST-FMR[17,18,19] spectra from all the devices of the six series of Pt/FM/CAP systems. The spectra provide fitted parameters that are used to quantify the magnitudes of the effective demagnetization field ($4\pi M_{eff}$), magnetic damping constant ($\alpha$), inhomogeneous linewidth-broadening ($\Delta H_0$) and the DL- and FL-SOT efficiencies ($\theta_{DL}$ and $\theta_{FL}$) for each device. Figures 3c−d illustrate the $4\pi M_{eff}$ of all the devices as a function of $1/t_{Co}^{eff}$ or $1/t_{Py}^{eff}$. The $4\pi M_{eff}$ of both Co and Py decreases with increasing $1/t_{FM}^{eff}$ regardless of the CAP. This is due to a presence of a strong $K_S$ that originates from the strong Pt 5$d$ – (Co,Fe) 3$d$ hybridization of the Pt/FM interface[20] and in part, from the (Co,Fe) 3$d$ – O 2$p$ hybridization[21] at the FM/oxide interface. The effective $K_S$ ($K_s^{eff}$) is estimated from the slope of the measured $4\pi M_{eff}$ vs $1/t_{FM}^{eff}$ using the relation $\partial(4\pi M_{eff})/\partial(1/t_{FM}^{eff}) = -(2K_S/M_S)$. The obtained $K_s^{eff}$ (Pt/Co/CAP) is as large as 1.0 erg/cm$^2$, 1.2 erg/cm$^2$ and 1.4 erg/cm$^2$ for the MgO-, HfO$_x$-, and TaN-CAP respectively. In addition, $K_s^{eff}$ (Pt/Py/CAP) is as large as 0.16 erg/cm$^2$, 0.22 erg/cm$^2$, and 0.29 erg/cm$^2$ for the same CAPs. Notably, the experimental $K_s^{eff}$ (Pt/Co/MgO/Ta) is very close to the previously reported[20] $K_S \approx$ 0.8–1.1 erg/cm$^2$ in Pt/Co/MgO.

We notice an apparent dependence of $K_s^{eff}$ on the CAP: $K_s^{eff}$ (MgO-CAP) < $K_s^{eff}$ (HfO$_x$-CAP) < $K_s^{eff}$ (TaN-CAP) for both Co and Py. The strength of PMA increases with a reduced formation of interfacial magnetic oxides. This is consistent with a recent work[22] in which the insertion of an ultrathin Hf layer between FeCoB (Py) and MgO layers reduces the oxidation of FeCoB (Py) and thereby enhances the PMA. The reduction of the PMA is also observed when there is an electric field induced O$^{2-}$ migration toward the Co layer in a Co/GdO$_x$ bilayer[23,24]. Based on these results, we cautiously assert that the interfacial oxidation of the FM compensates the surface magnetic anisotropy energy provided at the Pt/FM. The interfacial oxidation of FM deteriorates the PMA in the Pt/FM/CAP heterostructures.

The CAP material significantly influences the magnitudes of α and $\Delta H_0$ in Pt/Co/MgO/Ta (not in Pt/Py/MgO/Ta). Figures 4a−d present the obtained α and $\Delta H_0$ as a function of $1/t_{Co}^{eff}$ or $1/t_{Py}^{eff}$ where the thickness dependences of α and $\Delta H_0$ are quite different between Co and Py; both α and $\Delta H_0$ increase with decreasing $t_{FM}^{eff}$, but the increase is much more pronounced for the Pt/Co/MgO than for the Pt/Py/MgO. For the thinnest Co sample ($t_{Co}$ = 3 nm), the Pt/Co/MgO has α ≈ 0.06 which is at least three times as large as the α of the other CAPs (≈ 0.02). The α of different samples gets closer to each other as $t_{Co}^{eff}$ increases. Within the examined $t_{FM}$ ranges, α(Pt/Co/MgO) >> α(Pt/Co/HfO$_x$) ≈ α(Pt/Co/TaN); α(Pt/Py/MgO) ≈ α(Pt/Py/HfO$_x$) ≈ α(Pt/Py/TaN). Moreover, the α is proportional to $1/t_{eff}^2$ for the Pt/Co/MgO system, whereas it is quasi-linearly proportional to $1/t_{eff}$ for the rest of the system. A similar thickness dependence is observed with $\Delta H_0$. The results suggest that the Pt/Co/MgO system must have different origins for the enhancement of α and $\Delta H_0$.

There are two interfaces in Pt/FM/oxide systems which influence α and $\Delta H_0$. The Pt/FM interface contributes the enhancements in α and $\Delta H_0$ not only for Pt/Py/MgO/Ta but also for Pt/Co/CAP with the HfO$_x$ and TaN-CAPs, where both CAP materials negligibly contribute to the magnitudes of α and $\Delta H_0$. The major spin-dissipation channel is the spin-pumping (SP)[10] effect across the Pt/FM interface, which is clearly observed with the samples having a thin $t_{FM}^{eff}$. In contrast, the FM/oxide interface considerably influences the magnitudes of α and $\Delta H_0$ for the Pt/Co/MgO, implying that the Co/MgO interface develops an additional dissipation of spin angular momentum.

The XPS and ST-FMR results show that the existence of a significant CoO at the Co/MgO interface enhances α and $\Delta H_0$. The most probably mechanism for the additional relaxation is a two magnon scattering (TMS) process: the inhomogeneous contribution ($\Delta H_0$) becomes increasingly dominant for a thinner $t_{Co}^{eff}$. Below the Néel temperature ($T_N$), the interfacial magnetic oxides ($CoO$, $NiO$, and $Fe_2O_3$) become AF, because of the super-exchange interaction in which the spins of the 3d electrons in FM ions are ordered, yet oriented with respect to each other via the non-magnetic 2p electrons in $O^{2-}$. The AF-oxide opens additional relaxation pathways and spatially non-uniform spin dynamics during the magnetic precession via exchange coupling to the fluctuating spins of the AF oxide and the slow dragging of the AF-oxide domains.

To support the above-mentioned speculation, we need an independent evidence of AF order of the interfacial CoO. To investigate the AF order, the magnetic hysteresis loops of Pt/Co($t_{Co}$=3 nm)/MgO/Ta film were measured as a function of $T$, from 400 K to 10 K, using a SQUID-magnetometer. As shown in Fig. 5a, the loop is symmetric at RT (300 K). As we decrease $T$ down to cryogenic temperatures (e.g., 30 K) the loop is shifted ($\approx$ 250±120 Oe) along the axis of the applied in-plane magnetic field. This is an indicative of the exchange-bias effect between the Co and the interfacial CoO layers. The exchange bias field ($H_{ex}$) increases with a decrease in $T$ (see Fig. 5b), thus confirming the AF nature of the interfacial CoO layer. From the $T$-dependence of $H_{ex}$, the blocking temperature ($T_B$) of exchanged-biased Co/CoO is estimated to be approximately 150±50 K. The hysteresis loops of the Pt/Co($t_{Co}$=3 nm)/HfO$_x$ film were also measured, but no clear loop shift was observed within the measurement temperature (from 10 K to RT) and field step ($\approx$ 200 Oe). These observations are consistent with the XPS results in which the Pt/Co/HfO$_x$ system has less interfacial CoO than the Pt/Co/MgO/Ta.

A recent work[25] has demonstrated that Py thin films with an air-oxidized surface exhibit a significant increase in α near the magnetic phase transition ($T_N$) of the surface AF oxide layer. This was explained based on a magnetic fluctuation process. The surface AF layer develops magnon modes by enhancing the absorption of spin-angular momentum that is pumped from the oscillating FM layer. Spin fluctuation[26,27,28] theory suggests that the SP-efficiency is maximized at the $T_N$ of the AF-oxide, along with its long tail at higher $T$. This is because the magnetic susceptibility ($\chi$) of an AF-material is quite large even at its paramagnetic state (i.e., $\chi^{-1} \propto T+T_N$). Since $T_N$ is in general higher than $T_B$, we expect 150±50 K < $T_N$ (CoO at Co/MgO) < 300 K ($\approx T_N$ of bulk[29] CoO). Therefore, our CoO in the Pt/Co/MgO/Ta will exhibit a significant SP-effect across the Co/CoO interface even at RT where our ST-FMR measurement was done (see Fig. 1b). This can explain the substantial increase of damping in the Pt/Co/MgO/Ta.

Our XPS studies also revealed that the Pt/Py/MgO/Ta has a non-negligible amount of interfacial NiO and FeO, but their degrees of oxidization are much less than that of the CoO in Pt/Co/MgO/Ta. The interfacial (Ni,Fe)-oxide does not substantially enhance the α of the Pt/Py/MgO/Ta at RT (Fig. 4), presumably because the $T_N$ of the interfacial NiO and FeO is much lower than that of CoO [25]. The interfacial (Ni,Fe)-oxide would enhance the α as the $T$ approaches the $T_N$ [25,30]. For a more robust explanation of the linewidth broadening due to the FM/MgO interface, additional experiments are required such as a study of the FMR-linewidths as functions of a wide range of angles and frequency, as well as its $T$-dependence.

Finally, the effective $\theta_{DL}$ and $\theta_{FL}$ were obtained from the six different series of layer stack. In the ST-FMR measurement, the SOT-ratio efficiency ($\theta_{ratio}$) is calculated from the symmetric part ($V_S$) and the anti-symmetric part ($V_A$) of the FMR response using the expression $\theta_{ratio} = \frac{V_S}{V_A} \frac{e}{\hbar} 4\pi M_s t_{FM}^{eff} t_{Pt} \left(1 + \frac{4\pi M_{eff}}{H_{res}}\right)^{1/2}$, where $\hbar$ is the reduced Planck constant, $e$ is the electron charge, $H_{res}$ is the resonance field, and $t_{Pt}$ is the thickness of Pt (= 5 nm). According to the ST-FMR theory[14,18,31], the $V_S$ is proportional to $\tau_{DL}$ and the $V_A$ is due to the sum of $\tau_{FL}$ and the Oersted field torque ($\tau_{Oe}$). The relation between the effective $\theta_{DL}$ and $\theta_{FL}$ for a Pt/FM/CAP system is given as follows:

$$\frac{1}{\theta_{ratio}} = \frac{1}{\theta_{DL}}\left(1 + \frac{\hbar}{e}\frac{\theta_{FL}}{t_{Pt}}\frac{1}{4\pi M_s t_{FM}^{eff}}\right). \tag{1}$$

Then, the magnitude of $\theta_{DL}$ ($\theta_{FL}$) is calculated from the intercept (slope) of linear fit to the plot of $1/\theta_{ratio}$ vs $1/t_{FM}^{eff}$. This analysis is valid provided that (i) the $\theta_{DL}$ and $\theta_{FL}$ do not have a dependence on $t_{FM}^{eff}$ in the examined thickness range and (ii) no charge and spin currents flow through the CAP. Otherwise, the correct analysis should be done with the $V_S$ and $V_A$ measurement after a separate calibration of the microwave current through the device using a network analyzer[14].

Figures 5c–d present the results of $1/\theta_{ratio}$ vs $1/t_{Co}^{eff}$ or $1/t_{Py}^{eff}$ for the six series of layer stacks. Interestingly, regardless of the CAP, the slope for Co samples is negative whereas it is positive for Py samples. This indicates that the $\tau_{FL}$ is opposite to the $\tau_{Oe}$ (i.e., $\theta_{FL} < 0$) for the Pt/Co/CAP, whereas they are in the same direction for the Pt/Py/CAP ($\theta_{FL} > 0$). From the linear fits to Eq. (4), the $\theta_{DL}$ ($\theta_{FL}$) for the Pt/Co/CAP systems are estimated to be 0.089 (−0.012), 0.074 (−0.016), and 0.082 (−0.012) for the MgO-, HfOx-, and TaN-CAP respectively. In addition, the $\theta_{DL}$ ($\theta_{FL}$) for the Pt/Py/CAP system are evaluated to be 0.084 (0.015), 0.052 (0.008), and 0.072 (0.013) for MgO-, HfOx-, and TaN-CAP respectively. We conclude that the CAP materials in our stack configurations have a minor role in the determination of $\theta_{DL}$ and $\theta_{FL}$ within the experimental accuracy of measurements. Thus the results confirm that the SOTs originate mainly from the Pt/FM interface, and the contribution from the FM/CAP interface is very small. It is worth noting that the $\theta_{DL}$ of Pt/Co/CAP obtained by our work are slightly lower than the previously reported ones (≈ 0.12–0.14) by Ref. [19], whereas the magnitudes of $\theta_{DL}$ of our Pt/Py/CAP are larger than 0.05 from the same reference. Although both works have utilized the same measurement method (ST-FMR) from almost identical stacks, the estimated magnitudes could be different depending on whether the contribution of $\tau_{FL}$ is taken into consideration or not in the ST-FMR analysis. In addition, the values of our $\theta_{FL}$ (Pt/Py/CAP) and $\theta_{FL}$ (Pt/Co/CAP) are in agreement with Ref.

[31] and Ref. [18], respectively.

Presumably, the interfacial AF-oxide layer may lead a modification of $\theta_{DL}$ and $\theta_{FL}$ with the ultrathin FM ($t_{FM}$ < 2 nm). The FM layers used in this work are thick enough to absorb most of the spin-angular momentum transferred by the $J_S$ from the HM/FM interface, because the examined $t_{FM}$ ranges are larger than the penetration depth ($\lambda_p$) of the $J_S$ in the FM layers. The $\lambda_p$ is estimated to be 1–2 nm near the Fermi energy in 3d-FMs[32,33]; Ref. [33] suggests $\lambda_p$ ≈ 1.2 nm for Py, CoFeB, and Co. In $t_{FM}$ > $\lambda_p$, the role of the interfacial AF-oxide layer would be minimal in terms of determining the SOT-efficiencies. However, if $t_{FM}$ < $\lambda_p$, a portion of the $J_S$ will transmit to the spin-conducting AF-oxide layer[11,13], and thereby the resultant SOTs might be different. The $t_{FM}$ range from 0.5 to 2 nm is of critical importance because most of the SOT-driven magnetic switching of perpendicular magnetization have been demonstrated with this $t_{FM}$ range[6,7].

Several works have reported field-free[34,35,36] or beyond[37] SHE-driven perpendicular magnetization reversal in HM/FM/oxide multilayers, with some correlation to the FM-oxidization states[34,37]. The presence of AF-oxide, if the measurement temperature is below the $T_B$ of the AF-oxide, would lead to such field-free switching due to the symmetry breaking in the multilayer system. Alternatively, the interfacial transmission of $J_S$ into the AF-oxide and reflection back to FM might give rise to non-trivial SOT contribution to the FM layer (e.g., out-of-plane component of effective field or spin-polarization). Our results encourage further theoretical work which takes into account the interfacial AF-oxide in such HM/FM/oxide heterostructures (see Fig. 1b).

In summary, we illuminate an additional relaxation pathway of spin current in Pt/FM/CAP systems, namely the FM-oxide at the FM/CAP interface. The XPS showed the presence of interfacial FM-oxide, and ST-FMR demonstrated the deterioration of PMA associated with a formation of interfacial FM-oxides. The SOT-efficiencies ($\theta_{DL}$ and $\theta_{FL}$) were not substantially influenced by the CAP, confirming that both the DL- and FL-SOTs originate mainly from the Pt/FM interface. The interfacial oxide, especially CoO, significantly influences both α and $\Delta H_0$ indicating an extra magnetic damping, for instance, spin pumping process across the Co/CoO interface. This implies that the interfacial FM-oxide is a decent spin-current conductor, for instance, by incoherent magnon generation at the interfacial CoO, even above its magnetic ordering temperature ($T_N$). These results facilitate a better understanding of the interfacial-oxide contributions on the PMA, magnetic damping, and SOTs in Pt/FM/CAP systems.

## Acknowledgements

This work was supported by the National Research Council of Science & Technology (NST) grant (No. CAP-16-01-KIST) and the KIST Institutional Program (2E29410). K. -J. L. was supported by the National Research Foundation of Korea (NRF) [NRF-2017R1A2B2006119] and KU-KIST School Project. K. -J. L. acknowledges the KIST Institutional Program (Project No. 2V05750). Experimental data in Fig. 5a is obtained from KBSI SQUID VSM.

**Figure Captions**

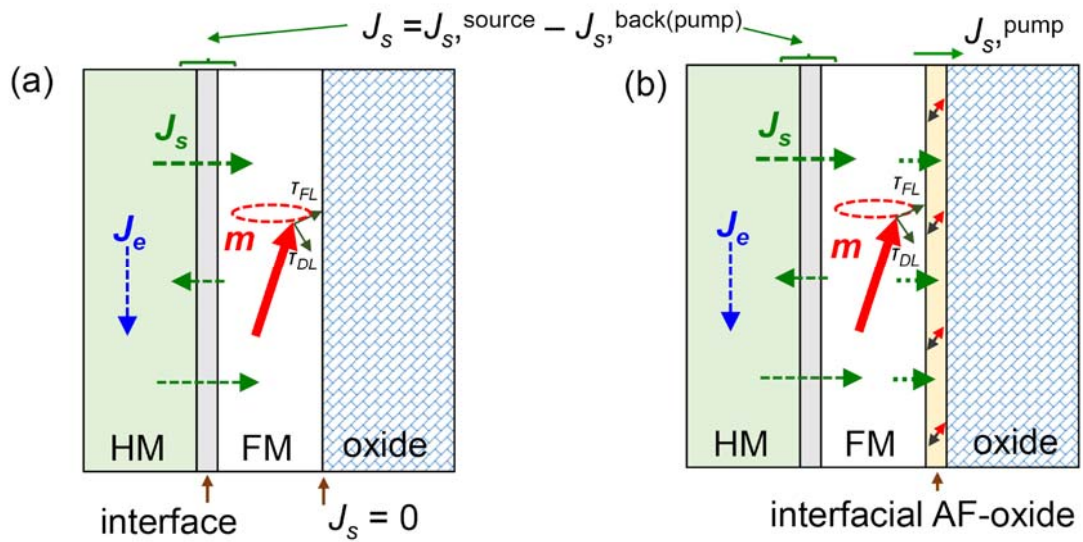

**Figure 1.** (a) Conventional view of magnetization dynamics induced by the $J_s$, and spin-pumping process in a HM/FM/oxide hetero-structure. The influence of the FM/oxide interface has been ignored. (b) A realistic view of the hetero-structure. There is a naturally formed AF-oxide layer at the interface between the FM and the oxide. A spin-fluctuation in the AF-oxide layer can provide an extra magnetic damping via the spin-pumping process across the FM/AF-oxide interface.

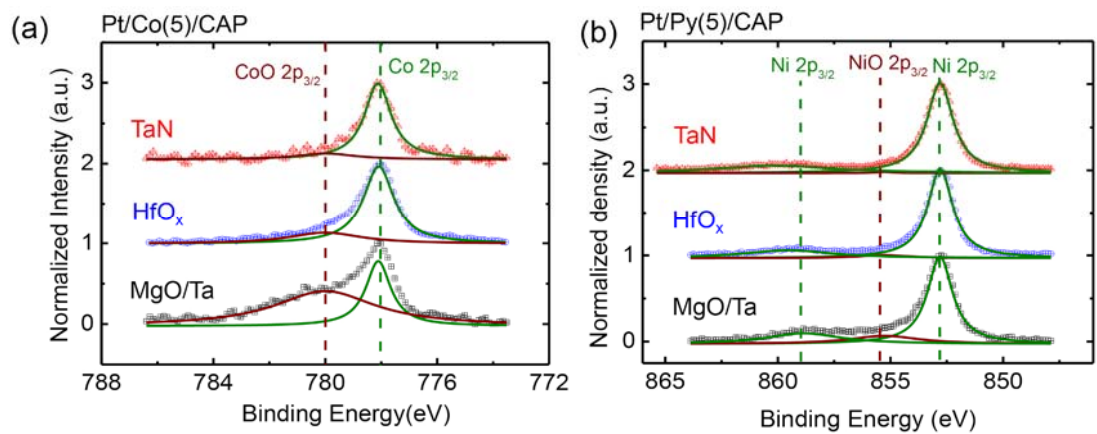

**Figure 2.** (a) XPS spectral of Co 2p3/2 from Pt/Co(5)/CAP samples. (b) XPS spectral of Ni 2p3/2 from Pt/Py(5)/CAP samples (nominal thickness in nm, and Py=$Ni_{79}Fe_{21}$). The CAP is either MgO/Ta (black), $HfO_x$ (blue) or TaN (red).

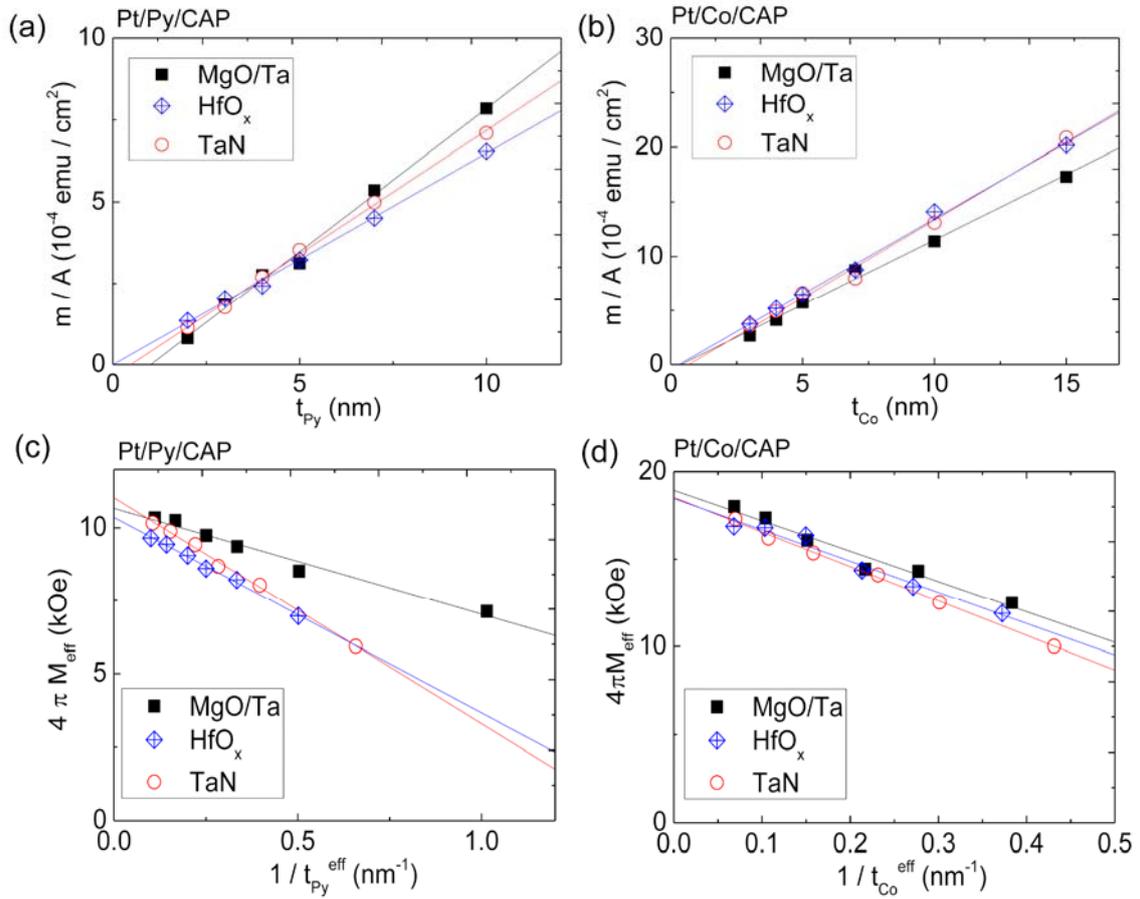

**Figure 3.** (a) Measured moment (per unit area) of un-patterned Pt/Py/CAP films as a function of $t_{Py}$. (b) Measured moment (per unit area) of un-patterned Pt/Co/CAP films as a function of $t_{Co}$. The linear fits provide saturation magnetization ($M_S$) and magnetic dead-layer thickness ($t^d$) for the MgO-, HfO$_x$- and TaN-CAP respectively. (c) Obtained $4\pi M_{eff}$ as a function $1/t_{Py}^{eff}$ from ST-FMR devices with Pt/Py/CAP multilayers. (d) Obtained $4\pi M_{eff}$ as a function $1/t_{Co}^{eff}$ from ST-FMR devices with Pt/Co/CAP multilayers.

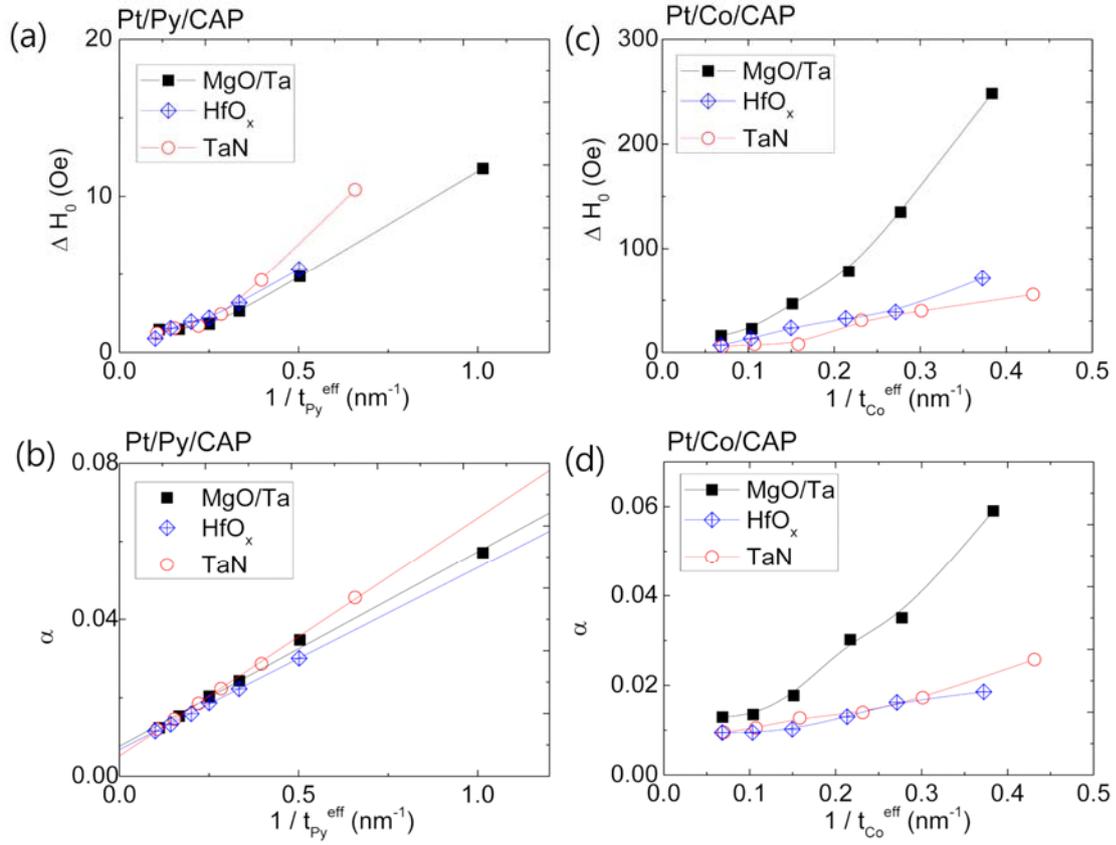

**Figure 4**. (a)-(b) Obtained α and Δ$H_0$ as functions of $1/t_{Py}^{eff}$ from ST-FMR devices with Pt/Py/CAP multilayers. (c)-(d) Obtained α and Δ$H_0$ as functions of $1/t_{Co}^{eff}$ from ST-FMR devices with Pt/Co/CAP multilayers.

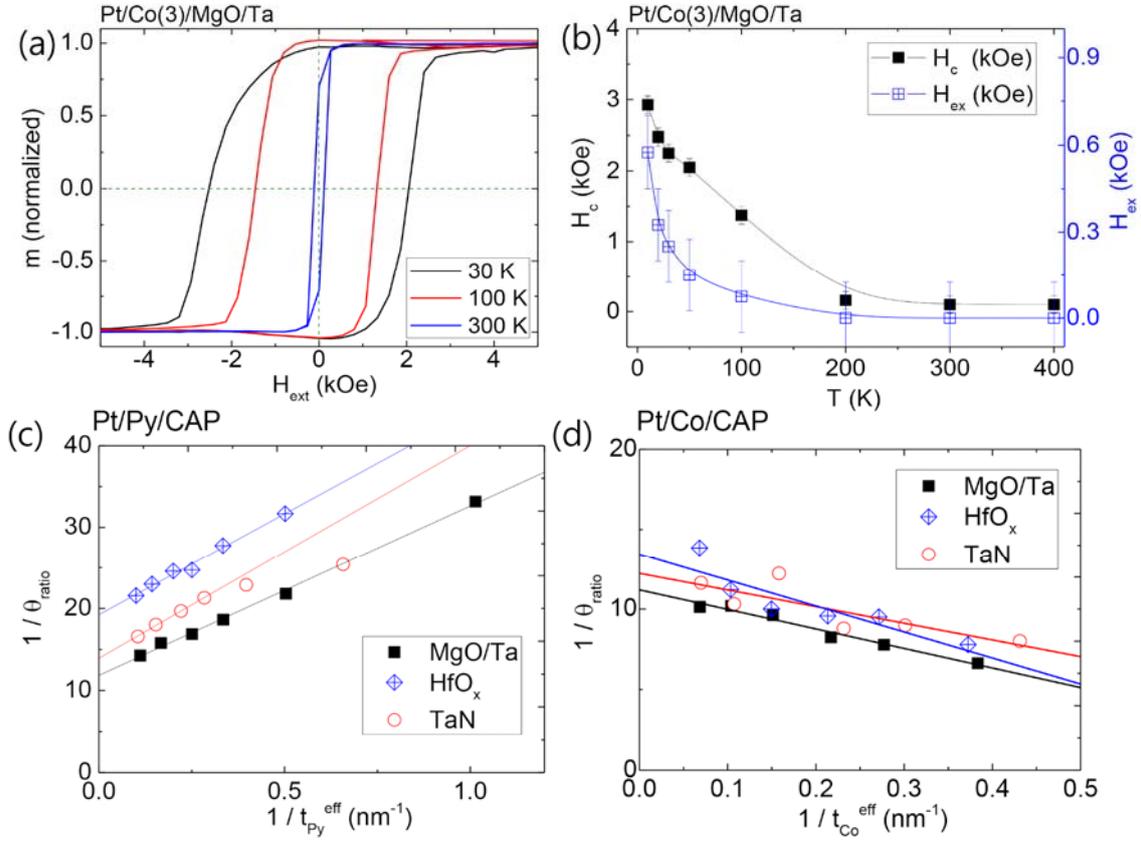

**Figure 5**. (a) Hysteresis loops of the Pt/Co($t_{Co}$=3 nm)/MgO/Ta film at 30, 100 and 300 K, measured by a SQUID-magnetometer. (b) Obtained coercive field ($H_c$) and loop center ($H_{ex}$) as a function of $T$. (c-d) Obtained $1/\theta_{ratio}$ as a function of (c) $1/t_{Py}^{eff}$ or (d) $1/t_{Co}^{eff}$ for the six series of layer stacks using ST-FMR measurement. The linear fit in the plot of $1/\theta_{ratio}$ vs $1/t_{FM}^{eff}$ provides the magnitude of $\theta_{DL}$ ($\theta_{FL}$), which can be calculated from the intercept (slope) respectively. The estimated $\theta_{DL}$ and $\theta_{FL}$ from all of the six layer stacks are summarized in Table. 1.

**Table**

|  |  | $t^d$ (nm) | $M_S$ (emu/cc) | $K_S$ (erg/cm$^2$) | $\theta_{DL}$ (×10$^{-2}$) | $\theta_{FL}$ (×10$^{-2}$) |
|---|---|---|---|---|---|---|
| Pt / Py / | MgO/Ta | 1.0 ± 0.2 | 870 ± 30 | 0.16 ± 0.01 | 8.4 ± 0.4 | 1.5 ± 0.1 |
|  | HfO$_x$ | 0.0 ± 0.2 | 650 ± 20 | 0.22 ± 0.01 | 5.2 ± 0.5 | 0.8 ± 0.1 |
|  | TaN | 0.5 ± 0.1 | 760 ± 20 | 0.29 ± 0.01 | 7.2 ± 0.4 | 1.3 ± 0.1 |
| Pt / Co / | MgO/Ta | 0.4 ± 0.3 | 1200 ± 50 | 1.0 ± 0.1 | 8.9 ± 1.0 | −1.2 ± 0.2 |
|  | HfO$_x$ | 0.3 ± 0.3 | 1400 ± 40 | 1.2 ± 0.1 | 7.4 ± 2.4 | −1.6 ± 0.5 |
|  | TaN | 0.7 ± 0.4 | 1430 ± 60 | 1.4 ± 0.1 | 8.2 ± 3.4 | −1.2 ± 0.5 |

**Table 1.** Summary of the measured magnetic dead layer ($t^d$), saturation magnetization ($M_S$), interfacial magnetic energy ($K_S$), SOT-efficiencies ($\theta_{DL}$ and $\theta_{FL}$) from the six different layer stacks.